# Mechanical properties and thermal conductivity of graphitic carbon nitride: A molecular dynamics study

Bohayra Mortazavi[1,2,*], Gianaurelio Cuniberti[2], Timon Rabczuk[1]

[1]*Institute of Structural Mechanics, Bauhaus-Universität Weimar, Marienstr. 15, D-99423 Weimar, Germany*

[2]*Institute for Materials Science and Max Bergman Center of Biomaterials, TU Dresden, 01062 Dresden, Germany*

Abstract

Graphitic carbon nitride ($g$-$C_3N_4$) nanosheets are among 2D attractive materials due to presenting unusual physicochemicalproperties.Nevertheless, no adequate information exists about theirmechanical and thermal properties. Therefore, we used classical molecular dynamicssimulations to explore thethermal conductivity and mechanical response of two main structures of single-layer triazine-based$g$-$C_3N_4$films.By performing uniaxial tensile modeling, we found remarkable elastic modulus of 320 and 210 GPa, and tensile strength of 44–47 GPa and 30 GPa for two different structures of $g$-$C_3N_4$sheets. Using equilibrium molecular dynamics simulations, the thermal conductivity of free-standing $g$-$C_3N_4$structures were also predicted to bearound 7.6 W/mK and 3.5W/mK. Our study suggests the $g$-$C_3N_4$films as exciting candidate for reinforcement of polymeric materials mechanical properties.

*Corresponding author (Bohayra Mortazavi):  bohayra.mortazavi@gmail.com

Tel: +49-176-68195567

Fax: +49 364 358 4511

# 1. Introduction

Two-dimensional (2D) materials that are consisting of few atomic layers have attracted tremendous attention due to presenting unique properties with promising prospects for extremely broad applications from nanoelectronics to aerospace structures. Thesignificance of 2D materials was primarily raised by the great success of graphene [1-3], with a unique combination of exceptionally high thermal [4], mechanical [5] and electrical properties [6]. The application of graphene in nanoelectronics is however limited because it does not present a bandgap and beingcategorized as a semi-metalic material. To surpass this limitation and open a bandgap in graphene, the common strategies involve complex physical or chemical modifications such as chemical doping [7-9]. Nevertheless, an alternative simpler rout is to use an inherent semiconducting materials like hexagonal boron-nitride(with bandgap of ~5.8 eV [10]) or $MoS_2$ sheets(1.8 eV bandgap[11]). However, due to chemical limitation, the desirable approach is yet to complement the electronic properties of the carbon-only graphite/graphene system that combines 2D atomic crystallinity and inherent semiconductivity [12]. To this aim, one possible candidate is graphitic carbon nitride (g-$C_3N_4$) which consists exclusively of covalently-linked, $sp^2$-hybridized carbon and nitrogen atoms [12, 13].Recently and for the first time, macroscopically large crystalline thin films of s-triazine-based, g-$C_3N_4$ (SgCN) films have been fabricated using an ionothermal, interfacial reaction [12]. Interestingly, experimental measurements and first principles density functional theory calculations suggest that SgCN has a bandgap of between 1.6 and 2.0 eV [12]. This short bandgap places the SgCN within the family of small bandgap semiconductors such as Si and GaAs, which proposesSgCN as an excellent candidate for nanoelectronic devices. We note that tri-triazine-based, g-$C_3N_4$ (TgCN) is also another form of graphitic carbon-nitride that has been experimentally fabricated and widely used for various applications. We remind thatin addition for application as 2D semiconducting materials in nanoelectronics, SgCN and TgCN filmshave great potential for energy conversion and storage, environmental applications including direct methanol fuel cells, catalysis, photocatalysis and $CO_2$ capture [14-17]. Graphitic carbon nitride are obviously "sustainable", as they are composed of carbon and nitrogen only, and can be prepared from simple precursors at a low cost [17].As a promising material to be used in a wide variety of applications, the comprehensive understanding of thermal and mechanical properties of g-$C_3N_4$ is also of a crucial importance. Due to difficulties of experimentalcharacterization of material properties at nanoscale, numerical and theoretical methods could be considered as an alternative approach [18]. To this purpose, we used classical molecular dynamics simulations to provide an overall image concerning the thermal conductivity and mechanical response of two main structures of triazine-based g-$C_3N_4$films.

## 2. Atomistic modeling

From theoretical point of view, there are several hypothetical phases of carbon nitride that naturally exist such as α, β, cubic, pseudocubic, and graphitic structures [17]. Among all these allotropes,g-$C_3N_4$ is understood to be the most stable structure under ambient conditions. Nevertheless, among different g-$C_3N_4$ structures, up to date, only s-triazine-based g-$C_3N_4$ (SgCN) has been synthesized in a form as that of graphite, consisting of macroscopic crystalline thin films [12]. On the other hand, tri-triazine-based, g-$C_3N_4$ (TgCN) is another form of graphitic carbon-nitridethat is widely used in chemical applications. Accordingly, we construct molecular models to evaluate the thermal and mechanical properties ofSgCN and TgCN sheets.In Fig. 1, atomic structure of SgCN and TgCNfilms areillustrated that are consisting of triazine units (ring of $C_3N_3$). In a SgCN film, a periodic array of single carbon vacancies exists while in a TgCN membrane three carbon atoms are removed in every repeating vacancy. We studied the tensile response and thermal conductivity of SgCN and TgCN films along armchair and zigzag directions (as shown in Fig. 1) in order to asses the intensity of anisothropic effects.

Molecular dynamics (MD) simulations in the present investigation were performed using LAMMPS(*Large-scale Atomic/Molecular Massively Parallel Simulator*) [19] package. The accuracy of molecular dynamics predictionsfor a particular structure is strongly dependent on the accuracy of utilized force fields for describing the interatomic forces. In this work, we used the Tersoff potential [20, 21] for introducing bonding interactions between carbon and nitrogen atoms. We usedthe set of parameters proposed by Lindsay and Broido [22] and Matsunaga *et al.* [23] for carbon and nitrogen atoms, respectively. Accordingly, the Tersoff potential parameters for carbon−nitrogen interactions were obtained using Eq. (1e) in Ref. [20]. It is worthy to note that the optimized Tersoff potential by Lindsay and Broido [22] could predict the phonon dispersion curves of graphite in closer agreement with experimental results than the original Tersoff potential. We also found that optimized Tersoff potential could predict the mechanical properties [24] and thermal conductivity [25] of pristine graphene in closer agreement with experiments than other common potential functions for the modeling of carbon atoms. In our previous work, we studied the mechanical response of nitrogen doped graphene. Interestingly, recent density functional theory study [26] confirms the accuracy of our predictions for mechanical properties of nitrogen doped graphene. We note that the time increment of all simulations was fixed at 0.5 fs.Based on the recent experimental findings [12], we assumed the thickness of 0.328 nm for SgCN and TgCN membranes.Moreover, in all MD simulations periodic boundary conditionswere applied in planar directions to minimize the finite length effects.

In this study, thermal conductivity of SgCN and TgCNfilms was evaluated using equilibrium molecular dynamics (EMD) method. To this aim, the structures were first relaxed to zero stress along the planar direction at room temperature using the Nosé-Hoover barostat and thermostat (NPT) method. The structures were then further equilibrated at constant volume and 300K using Berendsen thermostat method. At this point, the system is at accurate equilibrium conditions and the energy of the system is conserved by performing the constant energy (NVE) calculations. The EMD method relies on relating the ensemble average of the heat current auto-correlation function (HCACF) to the thermal conductivity $k$, via the Green-Kubo expression:

$$k_\alpha = \frac{1}{VK_BT^2}\int_0^\infty \langle J_\alpha(t)J_\alpha(0)\rangle\, dt \qquad (1)$$

where α denotes the three Cartesian coordinates, $K_B$ is the Boltzmann's constant, V and T are the volume and temperature of the system, respectively. The auto-correlation functions of the heat current $\langle J_\alpha(t)J_\alpha(0)\rangle$ can be calculated using the heat current $\vec{J(t)}$ as expressed by [19]:

$$\vec{J(t)} = \sum_i \left( e_i\vec{v_i} + \frac{1}{2}\sum_{i<j}(\vec{f_{ij}}\cdot(\vec{v_i}+\vec{v_j}))\vec{r_{ij}} \right) \qquad (2)$$

here, $e_i$ and $v_i$ are respectively the total energy and velocities of atom i, $f_{ij}$ and $r_{ij}$ are respectively the interatomic force and position vector between atoms i and j, respectively. By performing the constant energy simulations, the heat current values along armchair and zigzag directions were recorded to calculate the HCACFs. Several independent simulations were performed and the obtained HCACFs were averaged to calculate the effective thermal conductivity using Eq. 1.

The mechanical response of SgCN and TgCNsheetswas analyzed by performing uniaxial tensile simulations. Before applying the loading conditions, structure was relaxed to zero stress at room temperature using the NPT method. For the loading conditions, the periodic simulation box size along the loading direction was increased by a constant engineering strain rate of $1\times10^8$ s$^{-1}$ at every simulation time step. To ensure perfect uniaxial stress conditions, periodic simulation box along the sample width was altered using the NPT method. Virial stresses were calculated at each strain level to obtain engineering stress-strain response.

## 3. Results and discussions

In Fig. 2, the calculated thermal conductivity of SgCN and TgCN as a function of correlation time is shown. To ensure the size independency of calculated thermal conductivities, we performed the simulations for two samples with different numbers of atom. For each sample, the calculations were performed for ten different simulations with uncorrelated initial conditions. The results depicted in Fig. 2 are then the average of those along the armchair and zigzag directions. The EMD results for both samples reveal that at the correlation time of 5 ps, the thermal conductivity is well-converged. The obtained results clearly confirm the size indecency of estimated thermal conductivity for SgCN as well as TgCN films. Accordingly, based on the EMD approach, the thermal conductivity of free-standing SgCN is predicted to be 7.7±0.6 W/mK. In addition, our EMD results reveal that SgCN present around 12% higher thermal conductivity along the armchair direction in comparison with zigzag. On the other hand, we found a thermal conductivity of around 3.5±0.3 W/mK for suspended TgCN sheets. Our results for TgCN films suggest the independency of thermal conductivity to the chirality direction. The lower thermal conductivity of TgCN in comparison with SgCN is intrinsically because of increasing of phonon scattering rate due to the larger vacant area in TgCN films. Phonon scattering could also explain the isotropic thermal conductivity response in TgCN films. The predicted thermal conductivities for graphitic carbon-nitride structures are by two orders of magnitude lower than that of graphene sheets. We note that thermoelectric application of graphene is limited due its high thermal conduction properties. However, because of remarkably close atomic structures of graphene and g-$C_3N_4$ films, they could be considered as unique candidates for chemical integration. In this regard, fabrication of

graphene/g-C$_3$N$_4$hetero-structures could be considered as a promising approach for tuning the graphene thermal conductivity. This could enhance the thermoelectric figure of merit in favor to reach a high efficiency carbon-based thermoelectric material.

In Fig. 3, we plot the acquired uniaxial stress-strain responses of single-layer SgCN and TgCN films. To examine the intensity of loading direction effect on the mechanical properties, we studied the tensile response along armchair and zigzag loading directions. For SgCN films, our MD modeling yields elastic modulus of 320±5GPa and tensile strength of 44–47GPa at corresponding failure strain of 0.14–0.16.On another side, we found elastic modulus of 210±5 GPa and tensile strength of ≈30 GPa at rupture strain of ≈ 0.15 for TgCN sheets. The acquired MD results suggest the independency of elastic modulus on loading direction for both studied structures. On another hand, we foundthat SgCN films present slightly higher tensile strength along the zigzag direction than armchair whereas for the TgCN films the ultimate strength is shown to be independent of loading direction.The predicted mechanical properties by our MD study suggest the triazine-based, g-C$_3$N$_4$filmsas remarkably strong materials. It is worth mentioning that the predicted tensile strengths of 30 GPa and 44–47 GPa for graphitic carbonnitride filmsis by two orders of magnitude higher than that of the high strength steels and titanium alloys.Moreover, in comparison with defect-free graphene sheets, the elastic modulus and tensile strength of SgCN films are almost one third of those of graphene [5]. Presenting remarkably high mechanical properties, also recommend the SgCNand TgCN films as an excellent candidate to reinforce mechanical properties of polymeric materials [27]. Porous structure of SgCN and TgCN films leads to their lower stiffness and higher flexibility in comparison with graphene sheets. The high flexibility of additives could cause to reduce the stress concentration and postpone crack initiation in response. Therefore, in comparison with graphene flakes, it could be expected that nanocomposites made from graphitic carbon nitride films could present higher tensile strength responses. We note that due to presentence of saturation limits in enhancement of nanocomposites materials elastic response [28], the elastic stiffness of polymer nanocomposites made from SgCN or TgCN are expected to be close to that fabricated from same sized graphene flakes.

In Fig. 4, we illustrate the deformation process of a TgCN film along the armchair direction at different stages of loading. Our atomistic modeling suggests that during the loading condition the TgCN sheets extend uniformly and remain defect-free up to the tensile strength. We found that the tensile strength is a point that the first debonding occurs (Fig. 4a), which rapidly propagates (Fig. 4b and Fig. 4c) and leads to the specimen rupture (Fig. 4d). In this case, the initial debonding and subsequent rupture are found to occur at very close strain levels which accordingly suggest the brittle failure mechanism for TgCN nanosheets. As shown in Fig. 4d, during the crack growth in the sample, carbon-nitrogen chains form cross the sample that tend to keep the two sides of cracks connected.As it can be observed from Fig. 4d, the crack edges are mainly along the zigzag direction which reveals that debonding and crack growth happen by breaking of carbon-nitrogen bonds mostlyalong the armchair direction. In the case of SgCN films, we also found similar failure mechanism.

## 4. Summary

Triazine-based graphitic carbon nitridestructures are among attractive 2D-semiconducting material due to theirunusual physicochemicalproperties. Experimental studies have illustrated considerable prospects for their applications in nanoelectronics, energy conversion and storage, environmental applications, catalysis and photocatalysis. However, no adequate information exists about thermal conductivity and mechanical properties of these structures. Accordingly, we conductedatomistic simulations to study the mechanical properties and thermal conductivity of two main structures of graphitic carbon nitridestructures at room temperature. The Tersoff potential was used to introduce bonding interactions between carbon and nitrogen atoms in our molecular dynamics modeling.

Using equilibrium molecular dynamics simulations, the thermal conductivity of free-standing SgCN and TgCN were predicted to be 7.7±0.6 W/mK and 3.5 ±0.3 W/mK, respectively. Our results suggest that SgCN present around 12% higher thermal conductivity along the armchair direction in comparison with zigzag. On the other hand, by performing uniaxial tensile simulations, we found that pristineSgCN films can presentelastic modulus of 320±5 GPa and tensile strength of 44–47 GPa at failure strain of 0.14–0.16.Our MD results suggest that TgCN films can yield elastic modulus 210±5 GPa and tensile strength of around 30 GPa at rupture strain of around 0.15 indpendent of loading direction. Our results suggest the triazine-based graphitic carbon nitridestructuresas remarkably strong and flexible 2D materialswell suited for reinforcement of mechanical properties of polymeric materials.

**References**

[1] K. S. Novoselov, A. K. Geim, S. V. Morozov, D. Jiang, Y. Zhang and S. V. Dubonos, et al., Electric field effect in atomically thin carbon films, Science, 2004, 306, 666–669.

[2] K. S. Novoselov, D. Jiang, F. Schedin, T. J. Booth, V. V. Khotkevich and S. V. Morozov, et al., Two-dimensional atomic crystals, Proc. Natl. Acad. Sci. U. S. A., 2005, 102, 10451–10453.

[3] A. K. Geim , K. S. Novoselov,The rise of graphene. Nat Mater., 2007, 6, 183–191.

[4] S. Ghosh, W. Bao, D. L. Nika, S. Subrina, E. P. Pokatilov and C. N. Lau, et al., Dimensional crossover of thermal transport in few-layer graphene, Nat. Mater., 2010, 9, 555–558.

[5]C. Lee, X. Wei, J. W. Kysar and J. Hone, Measurement of the elastic properties and intrinsic strength of monolayer graphene, Science, 2008, 321, 385–388.

[6] J. R. Williams, L. DiCarlo and C. M. Marcus, Quantum Hall effect in a gate-controlled pn junction of graphene, Science, 2007, 317, 638–641.

[7] T.B. Martins, R.H. Miwa, A.J.R. da Silva, A. Fazzio, Electronic and transport properties of boron-doped graphene nanoribbons, Phys. Rev. Lett., 2007, 98, 196803.

[8] A. Lherbie, R. X. Blasé, Y. Niquet, F. Triozon, S. Roche, Charge transport in chemically doped 2D graphene, Phys. Rev. Lett., 2008, 101, 036808.

[9] L. Ci, L. Song, C. Jin, D. Jariwala, D. Wu, Y. Li, A. Srivastava, et al., Atomic layers of hybridized boron nitride and graphene domains, Nat. Mater., 2010, 9, 430-435.

[10] K. Watanabe, T. Taniguchi, H. Kanda, Direct-bandgap properties and evidence for ultraviolet lasing of hexagonal boron nitride single crystal, Nat. Mater, 2004, 3, 404 – 409.

[11] B. Radisavljevic, A. Radenovic, J. Brivio, V. Giacometti, A. Kis, Single-layer $MoS_2$ transistors, Nat. Nanotech., 2011, 6, 147-150.

[12] G. Algara-Siller, N. Severin, S.Y. Chong, T. Björkman, R. G. Palgrave, A. Laybourn, *et al.* Triazine-based, graphitic carbon nitride: a two-dimensional semiconductor, Angew. Chem., 2014, 126, 1-6.

[13] D. M. Teter, R. J. Hemley, Low-Compressibility Carbon Nitrides, Science, 1996, 271, 53–55.

[14] X. Wang, K. Maeda, A. Thomas, K. Takanabe, G. Xin, J. M. Carlsson, K. Domen and M. Antonietti, A metal-free polymeric photocatalyst for hydrogen production from water under visible light. Nat. Mater., 2008, 8, 76–80.

[15] Y. Zheng, Y. Jiao, J. Chen, J. Liu, J. Liang, A. Du, et al., Nanoporous Graphitic-$C_3N_4$ Carbon Metal-Free Electrocatalysts for Highly Efficient Oxygen Reduction. J. Am. Chem. Soc., 2011, 133, 20116–20119.

[16] S. M. Lyth, Y. Nabae, N. M. Islam, S. Kuroki, M. Kakimoto and S. Miyata, Electrochemical Oxygen Reduction Activity of Carbon Nitride Supported on Carbon Black, J. Electrochem. Soc.2011, 158, B194–B201.

[17] Y. Zheng, J. Liu, J. Liang, M. Jaroniec and S. Z. Qiao, Graphitic carbon nitride materials: controllable synthesis and applications in fuel cells and photocatalysis. Energy Environ. Sci., 2012, 5, 6717

[18] A. A. Khatibi, B. Mortazavi, A study on the nanoindentation behaviour of single crystal silicon using hybrid MD-FE method, Adv. Mat. Res. 208, 32, 259-262.

[19] S. Plimpton,Fast parallel algorithms for short-range molecular dynamics. J. Comp.Phys.1995, 117, 1-19.

[20] J. Tersoff, New empirical approach for the structure and energy of covalent systems, Phys. Rev. B, 1988, 37, 6991–7000

[21] J. Tersoff, Empirical interatomic potential for carbon, with applications to amorphous carbon, Phys. Rev. Lett., 1988, 61, 2879–2882.

[22] L. Lindsay and D. A. Broido, Optimized Tersoff and Brenner empirical potential parameters for lattice dynamics and phonon thermal transport in carbon nanotubes and graphene, Phys. Rev. B, 2010, 82, 205441.

[23] K. Matsunaga, C.Fisher, H.Matsubara,Tersoff potential parameters for simulating cubic boron carbonitrides. Jpn. J. Appl. Phys., 2000, 39, 48-51.

[24] B. Mortazavi, G. Cuniberti, Atomistic modeling of mechanical properties of polycrystalline graphene, Nanotechnology, 2014, 25, 21570.

[25] B. Mortazavi, S. Azi, Thermal conductivity and tensile response of defective graphene: A molecular dynamics study, Carbon, 2013, 63, 460-470.

[26] K.Z. Milowska, M. Woinska, M. Wierzbowska, Contrasting Elastic Properties of Heavily B-and N-doped Graphene with Random Impurity Distributions Including Aggregates, J, Phys. Chem. C, 2013, 117, 20229-20235.

[27] B. Mortazavi, F. Hassouna, A. Laachachi, A. Rajabpour, S. Ahzi, D. Chapron, et al.,Experimental and multiscale modeling of thermal conductivity and elastic properties of PLA/expanded graphite polymer nanocomposites, Thermochimica Acta, 2013, 552, 106-113.

[28] B. Mortazavi, J. Bardon, S. Ahzi, Interphase effect on the elastic and thermal conductivity response of polymer nanocomposite materials: 3D finite element study. Comput. Mat. Sci. 69, 2013, 100-106.

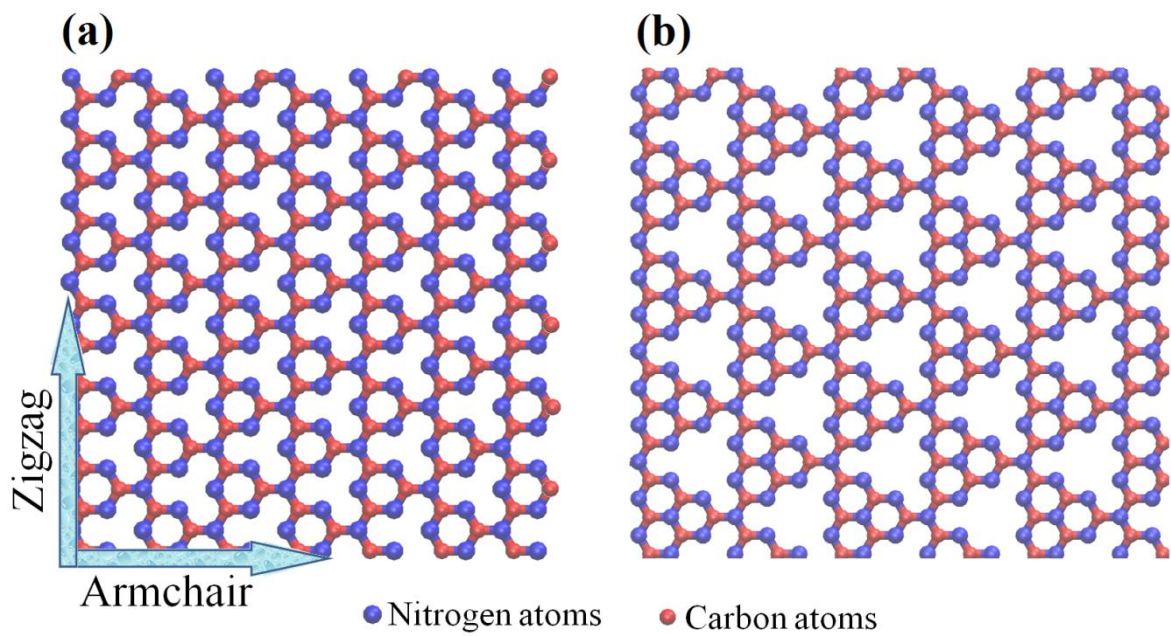

Fig.1-Atomic structure of (a) s-triazine-basedg-$C_3N_4$(SgCN) and (b) tri-triazine-basedg-$C_3N_4$ (TgCN).

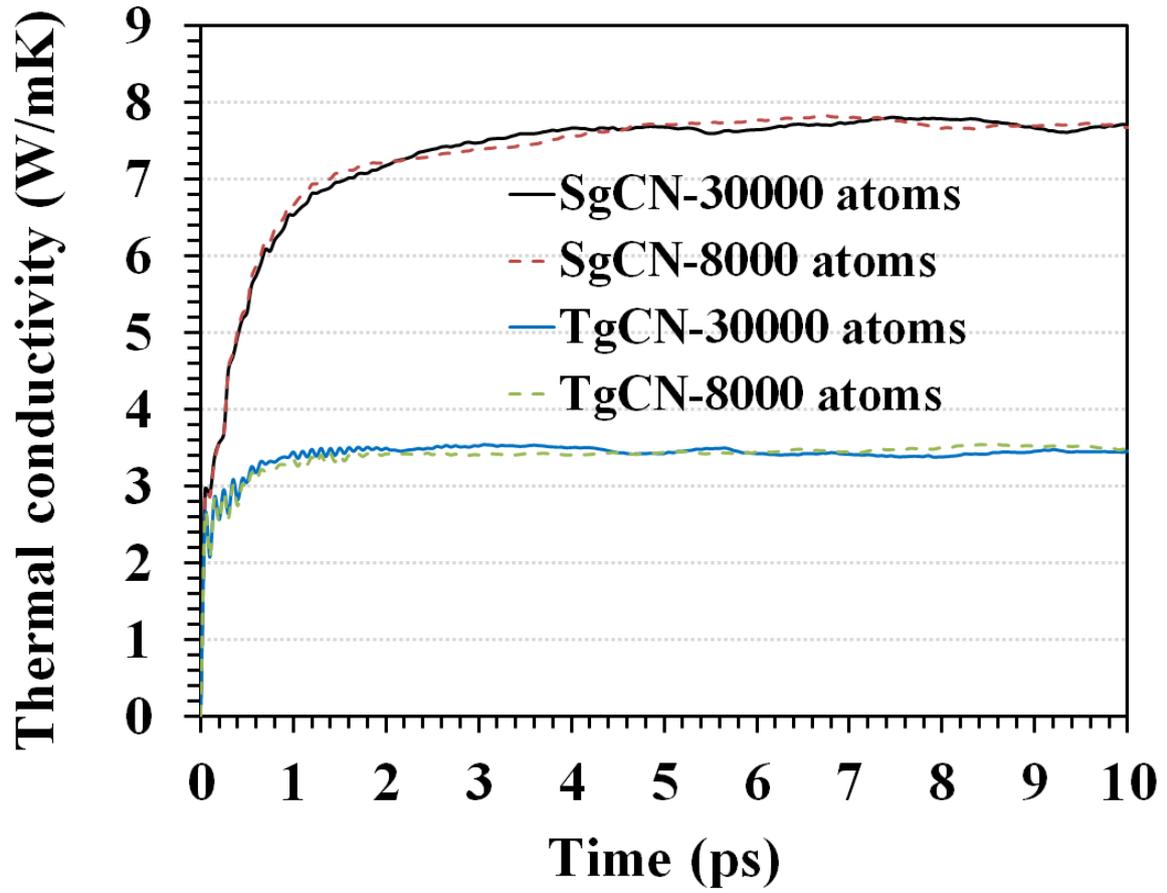

Fig. 2-Calculated thermal conductivity of single-layer SgCN and TgCN sheets as a function of correlation time for two systems with different numbers of atom.

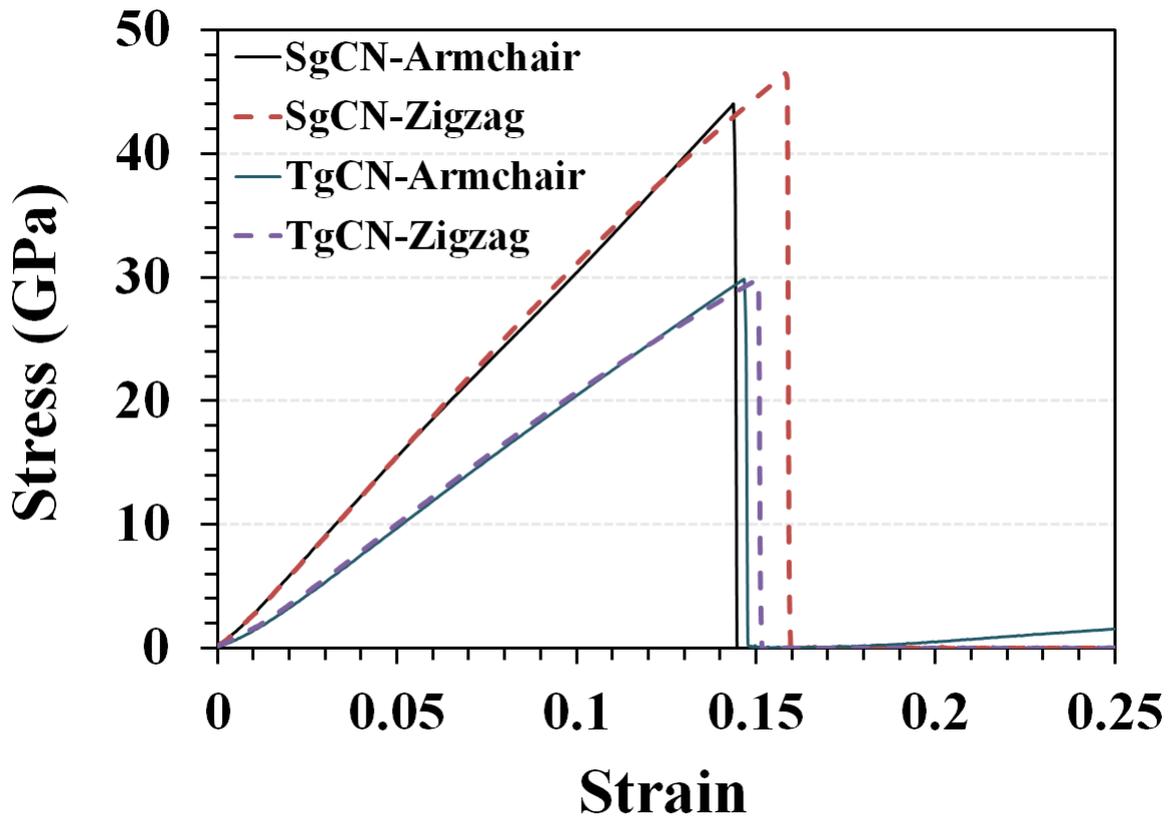

Fig. 3- Effect of loading direction on the stress-strain response of TgCN and SgCN films.

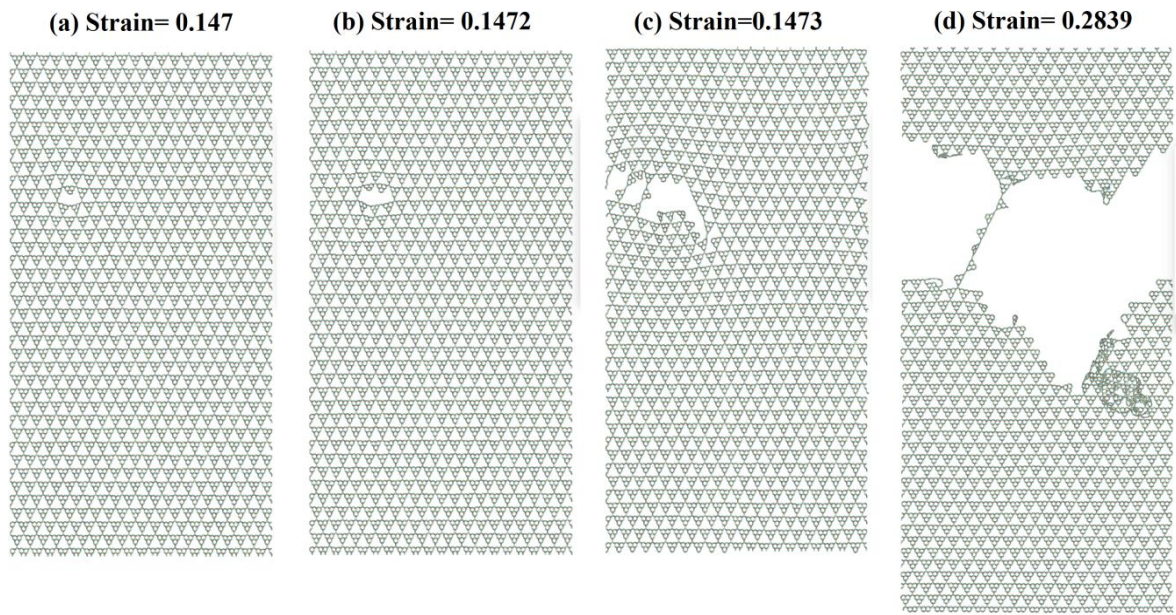

Fig. 4-Failure process of a TgCN membranealong the armchair direction at different engineering strain levels.